\pdfoutput=1
\documentclass[11pt]{article}

\usepackage[final]{acl}

\usepackage{times}
\usepackage{latexsym}

\usepackage{amsmath}
\usepackage{bbding}
\usepackage{lipsum}
\usepackage{listings}
\usepackage[ruled,vlined]{algorithm2e}
\usepackage{multirow}
\usepackage{graphicx}
\usepackage{cleveref}
\usepackage{hyperref}
\usepackage{subcaption}
\usepackage{pifont}
\usepackage[most]{tcolorbox}
\usepackage{xcolor}

\definecolor{colframecolor}{RGB}{55,98,175}   % frame color (边框色)
\definecolor{colbackcolor}{RGB}{218,227,243}  % background color (背景色)

\usepackage{booktabs}
\usepackage{makecell} % 前导导言区需要引入

\usepackage{fontawesome}
\usepackage[T1]{fontenc}
\usepackage[utf8]{inputenc}

\usepackage{microtype}

\usepackage{inconsolata}

\usepackage{graphicx}

\title{CTR-Guided Generative Query Suggestion in Conversational Search}

\author{
 \textbf{Erxue Min\textsuperscript{1}$^{\ast}$},
 \textbf{Hsiu-Yuan Huang\textsuperscript{1,2}$^{\ast}$},
 \textbf{Xihong Yang\textsuperscript{1}$^{\ast}$},
 \textbf{Min Yang\textsuperscript{1}},\\
 \textbf{Xin Jia\textsuperscript{1}$^{\dagger}$},
 \textbf{Yunfang Wu\textsuperscript{2}},
 \textbf{Hengyi Cai\textsuperscript{1}},
 \textbf{Junfeng Wang\textsuperscript{1}},
 \textbf{Shuaiqiang Wang\textsuperscript{1}},
 \textbf{Dawei Yin\textsuperscript{1}}
\\
\\
 \textsuperscript{1}Baidu Inc,
 \textsuperscript{2}Peking University,
\\
{$^{\ast}$: Equal contribution. $^{\dagger}$: Corresponding author.}
}

\begin{document}
\maketitle

\begin{abstract}
Generating effective query suggestions in conversational search requires aligning model outputs with user preferences, which is challenging due to sparse and noisy click signals. We propose \textbf{GQS}, a generative framework that integrates click modeling and preference optimization to enhance real-world user engagement.
GQS consists of three key components: (1) a \textit{Multi-Source CTR Modeling} module that captures diverse contextual signals to estimate fine-grained click-through rates; (2) a \textit{Diversity-Aware Preference Alignment} strategy using CTR-weighted Direct Preference Optimization (DPO), which balances relevance and semantic diversity; and (3) a \textit{CTR-Calibrated Iterative Optimization} process that jointly refines the CTR and generation models across training rounds.
Experiments on two real-world tasks demonstrate that GQS outperforms strong baselines in CTR, relevance, and diversity. %Further analysis shows that dynamic reward calibration and diversity-aware training are critical to aligning generation with user behavior.
\end{abstract}

\section{Introduction}

\begin{figure}[t]
    \centering
    \includegraphics[width=\columnwidth]{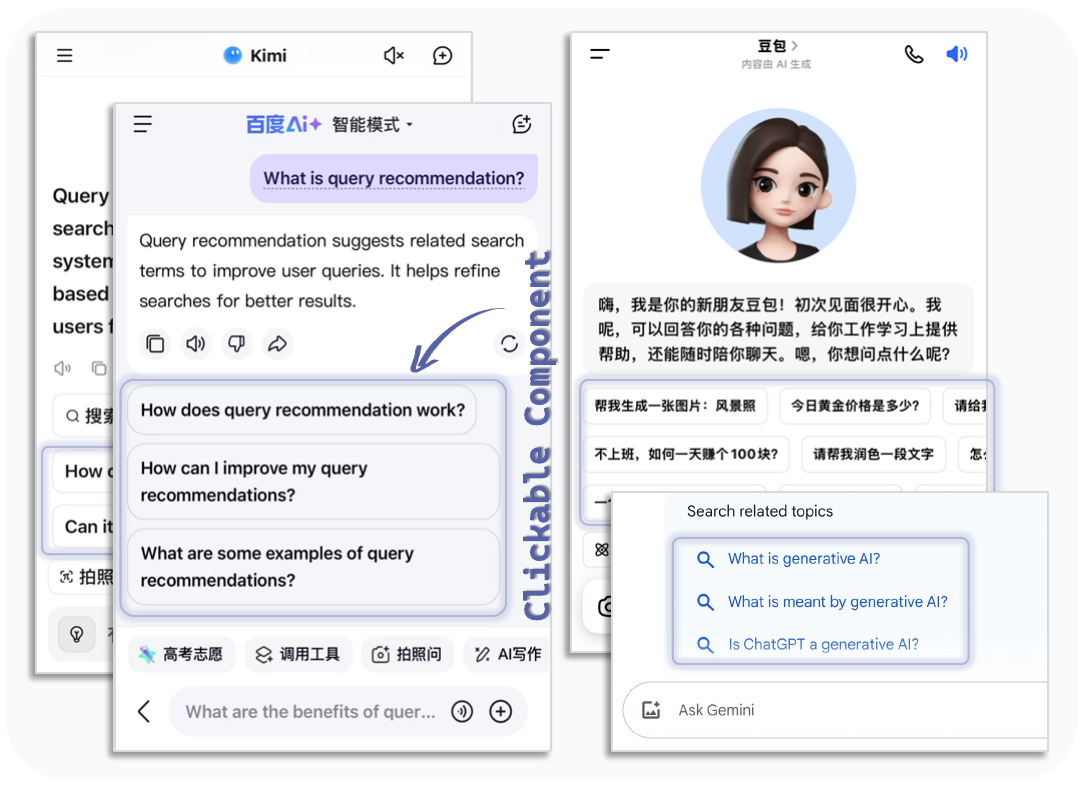} % 
    \caption{Examples of AI assistant interfaces where query suggestions are presented as clickable elements.
    }
    \label{fig:teaser}
\end{figure}

% Conversational AI assistants have emerged as a new paradigm for search, 
Conversational search systems such as AI assistants have emerged as a new paradigm for search,
where users interact through natural language queries and receive not only direct answers but also proactive query suggestions. As shown in Figure~\ref{fig:teaser}, these suggestions are typically presented as clickable elements in the interaction interface, helping users refine or extend their queries with minimal effort.

Recent work on query suggestion has explored using general-purpose Large Language Models (LLMs) as the backbone, leveraging their internal knowledge to alleviate cold-start limitations and generate suggestions in conversational search~\cite{10.1145/3583780.3614949,ZeroShot-CQGen-4ConversationalSearch,MiningExploratoryQueries_Dou,MultiTurnClarificationDou,2024MultimodalQuerySuggestion}. While these models produce fluent and plausible suggestions, they lack grounding in real user search behavior and often fail to align with actual user preferences. Retrieval-augmented generation (RAG) frameworks~\cite{bacciu2024generatingqueryrecommendationsllms,shen2024enhancing,wang2024richragcraftingrichresponses} incorporate external search-related knowledge to bridge this gap, but still depend on LLMs’ retrieval and summarization abilities rather than genuine preference alignment. 
% In conversational AI assistant scenarios, user clicks offer the most direct signal of preference. Yet, how can these signals be effectively leveraged to guide query suggestion?
In conversational search systems, user clicks provide one of the most direct signals of user preference. However, how to effectively leverage such signals to guide query suggestion remains an open challenge.

% As a result, these methods generate outputs that, while coherent, often fail to meet real user needs due to the absence of feedback calibration based on authentic interaction signals.

% We identify three key challenges in building effective click-preference-aligned query suggestion systems for conversational AI assistants:
We identify three key challenges in building effective click-preference-aligned Generative Query Suggestion (GQS) framework:
(1) Real-world click data is inherently noisy and biased, making it unsuitable for out-of-the-box use \cite{10.1145/3681785,10.1145/3589335.3651576,islam2025correctingpositionbiaslearning}. This calls for reliable click-through rate (CTR) modeling to calibrate and denoise click signals before applying them for preference alignment.
(2) Directly optimizing for CTR often reduces the semantic diversity of suggestions, resulting in repetitive or narrowly focused outputs \cite{peng2023reconcilingaccuracydiversitytradeoffrecommendations, kirk2024understandingeffectsrlhfllm, 10.1145/3626772.3657736, zhao2025echochamberrlposttraining}. Balancing CTR alignment with diversity preservation is essential to sustain user engagement.
(3) Collecting sufficient online feedback for preference optimization is time-consuming and costly, underscoring the importance of fully exploiting existing offline click data through efficient iterative improvement strategies \cite{chen2023opportunitieschallengesofflinereinforcement,kang2023rewardofflinepreferenceguidedpolicy,10.1145/3626772.3657736}.

In this paper, we propose a novel generative framework for AI assistant query suggestion that addresses these challenges through three key components:
(1) a \textbf{\textit{Multi-Source CTR Modeling}} strategy that integrates multiple signals to build a robust CTR predictor for reliable click signal estimation.
(2) a \textbf{\textit{Diversity-Aware Click Preference Alignment}} method that jointly optimizes CTR alignment and semantic diversity using Direct Preference Optimization (DPO) over diversity-enhanced preference pairs.
(3) a \textbf{\textit{CTR-Calibrated Iterative Optimization}} process that applies importance sampling to effectively reuse offline click data for iterative preference alignment, enabling continuous improvement without relying on costly online feedback collection.

Our experiments on industrial-scale AI assistant datasets demonstrate substantial improvements in CTR and diversity metrics, validating the effectiveness of our approach. To summarize, our contributions are:
\begin{itemize}
\item We propose an end-to-end Generative Query Suggestion (GQS) framework tailored for AI assistant query suggestion.
\item We construct Multi-Source CTR Modeling for Diversity-Aware Click Preference Alignment to align generation with real user preferences, and propose a CTR-Calibrated Iterative Optimization approach for efficient improvement.
\item We validate our method through large-scale A/B online experiments, showing significant improvements in user engagement and suggestion quality.
\end{itemize}

% \section{Preliminaries}
\section{Background}
\label{sec:prelim}
We formulate query suggestion in conversational search as a conditional generation task, where an LLM generates novel queries tailored to the user’s evolving intent instead of retrieving queries from a fixed set. This enables proactive exploration beyond explicit user input, improving both experience and search coverage.

To maximize the quality and relevance of generated suggestions, our prompt design explicitly incorporates multiple contextual sources as side information, as illustrated in Appendix~\ref{sec:prompt}:  
\textbf{(1)} \textit{Current user query} $q_{u(\tau)}$, representing the user's immediate information need at conversation turn $\tau$;  
\textbf{(2)} \textit{Current assistant response} $r_{(\tau)}$, showing how the system interprets and addresses the current query;  
\textbf{(3)} \textit{Conversation history} $h_{(\tau)}$, consisting of all prior user–assistant turns before $\tau$, providing broader dialog context;  
\textbf{(4)} \textit{User profile features} $u$, including age, interests, or behavioral tags, to encode user background and preferences;  
\textbf{(5)} \textit{Co-occurring queries} $\mathcal{C}_{(\tau)}$, mined from similar historical conversations as external user behaviour references (see Appendix~\ref{sec:coo_con} for construction details).

Formally, we represent the context as $\mathcal{X}_{(\tau)} = \{{q_{u(\tau)}, r_{(\tau)}, h_{(\tau)}, u, \mathcal{C}_{(\tau)}}$\}. The LLM is prompted to jointly generate a set of candidate queries $\hat{Q}^{(t)}_{(\tau)} = \{\hat{q}_1, \hat{q}_2, \dots, \hat{q}_N\}$ from the conditional distribution:
\begin{equation}
\hat{Q}^{(t)}_{(\tau)} \sim \pi_{\theta}^{(t)}(q_1, \dots, q_N \mid \mathcal{X}_{(\tau)}),
\end{equation}
where $\pi_{\theta}^{(t)}$ denotes the search preference-aligned LLM under the current optimization turn $t$, and $\hat{Q}^{(t)}_{(\tau)}$ is the resulting set of $n$ candidate suggestions.
By explicitly emphasizing current user query alongside above side information as content $\mathcal{X}_{(\tau)}$, this prompt structure guides the model to generate suggestions $\hat{Q}^{(t)}_{(\tau)}$ related to user search intent while maintaining holistic awareness. 
 
%This single-pass generation design supports explicit control over suggestion diversity, minimizes computational overhead, and enables effective modeling of inter-query dependencies—forming the foundation of our CTR-driven query recommendation pipeline.

%To further enhance               the generated suggestion candidate set $\hat{Q}^{(t)}_{(\tau)}$, we structure the LLM optimization process $t$ into two stages: an initialization stage at $t=0$ for \textit{Diversity-Aware Click Preference Alignment}, and a refinement stage for $t>0$ via \textit{CTR-Calibrated Iterative Optimization}.

\begin{figure*}[t]
    \centering
    \includegraphics[width=1\linewidth]{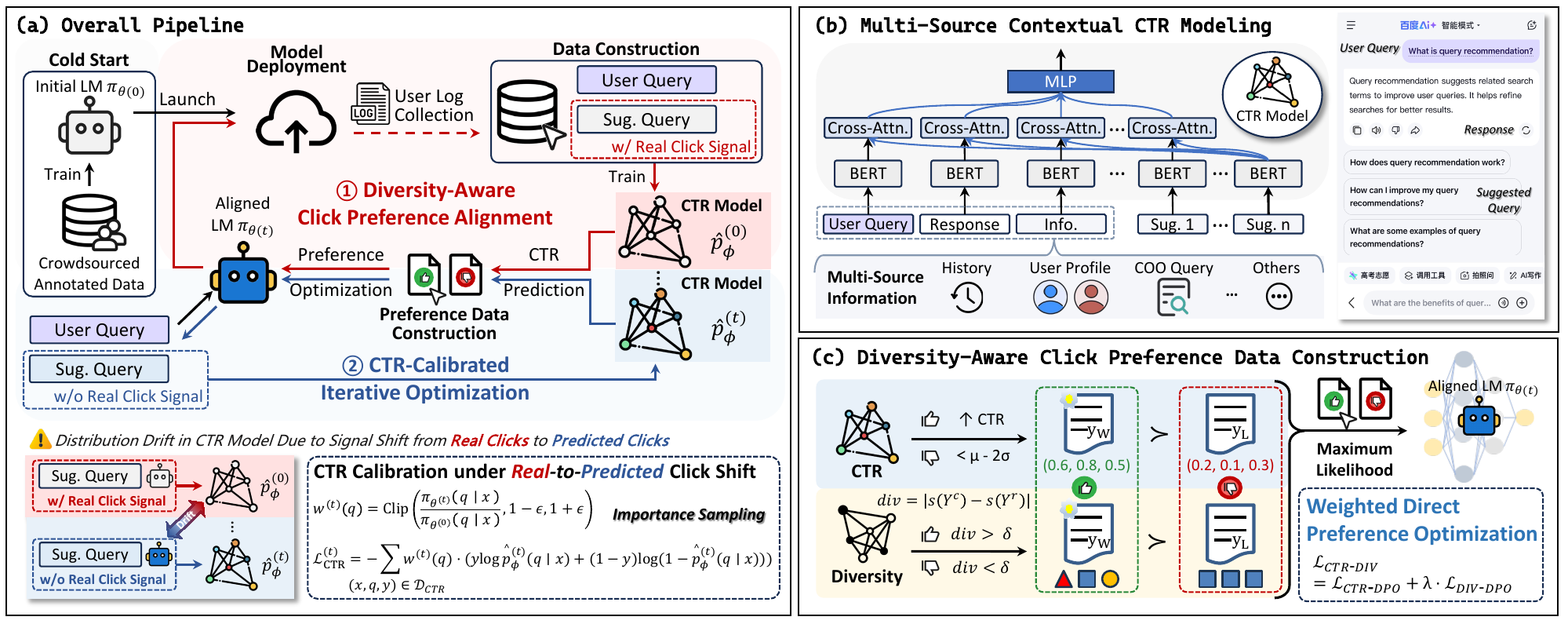} % 
    \caption{An overview of our proposed method. (a) The overall pipeline of the framework. (b) Multi-Source contextual Modeling for reliable CTR estimation. (c) Diversity-Aware Click Preference Data Construction for balancing CTR alignment and diversity.
    }
    \label{fig:main}
\end{figure*}

\section{Methodology}
\label{sec:method}
In this section, we introduce our framework for CTR-driven generative query suggestion in conversational AI assistent search. The method consists of three major components:
(1) a multi-source contextual CTR modeling to obtain click prefence signals;
(2) CTR-weighted Direct Preference Optimization (DPO) with diversity regularization for click preference alignment;
(3) an iterative training scheme with calibrated CTR correction to ensure efficient policy improvement.

\subsection{Multi-Source Contextual CTR Modeling}

User click signals extracted from logs offer a natural form of preference supervision, but are inherently noisy and biased (e.g., due to position effects), making them unsuitable for direct use. A central challenge in aligning a GQS LLM with real user preferences lies in accurately estimating the click-through rate (CTR) of each generated query under diverse conversational contexts.

To address this, we propose a multi-source CTR prediction model that:
(1) encodes heterogeneous contextual signals,
(2) uses cross-attention to model the interaction between each context and the target query, and
(3) integrates these signals for final CTR estimation, as illustrated in Figure \ref{fig:main}(b).

Let $q_n$ denote the $n$-th generated query suggestion, where $n \in \{1, 2, \dots, N\}$. Let $\mathcal{X} = \{x_1, x_2, \dots, x_K\}$ represent a set of contextual sources, including the user’s current query, AI response, dialog summary, user profile, co-occurring queries, and prior generated queries. Each $x_k \in \mathcal{X}$ is encoded using a shared BERT encoder:
\[
\mathbf{H}_k = \text{BERT}(x_k), \quad \mathbf{H}_n = \text{BERT}(q_n).
\]

To model the relationship between $q_n$ and each context $x_k$, we apply single-head cross-attention:
\[
\mathbf{A}_k = \text{softmax}\left( \frac{\mathbf{H}_n \mathbf{W}_Q (\mathbf{H}_k \mathbf{W}_K)^\top}{\sqrt{d}} \right)(\mathbf{H}_k \mathbf{W}_V),
\]
\[
\mathbf{e}_k = \text{Pool}(\mathbf{A}_k),
\]
where $\mathbf{W}_Q$, $\mathbf{W}_K$, and $\mathbf{W}_V$ are learnable projection matrices, and $\text{Pool}(\cdot)$ denotes mean pooling over attended tokens to produce fixed-length representations.
To account for position bias in the generated query list, we incorporate a learnable position embedding $\mathbf{e}_{\text{pos}} = \text{Embed}(p)$, where $p$ is the position index of $q_n$.

The final CTR prediction for $q_n$ is computed by concatenating all contextual embeddings and feeding them into a multi-layer perceptron:
\[
\hat{y}_n = \sigma\left( \text{MLP}\left( \left[ \mathbf{e}_1; \dots; \mathbf{e}_K; \mathbf{e}_{\text{pos}} \right] \right) \right).
\]
The model is trained using binary click labels as supervision, optimizing a standard binary classification loss.
This attention-based architecture enables fine-grained modeling of how each type of context contributes to click likelihood, supporting CTR-informed preference data construction in the next stage.

\subsection{CTR-Driven Preference Optimization for Query Generation}
\label{sec:CTR-DrivenPO}
To align the GQS model with real user preferences, we formulate training as a preference-based learning problem. Rather than directly maximizing predicted CTR which may be poorly calibrated, we aim to optimize for response-level preference rankings inferred from CTR scores. We build upon DPO~\cite{NEURIPS2023_DPO}, extending it with CTR-weighted importance and an auxiliary diversity objective.

\paragraph{Step 1: Response Scoring and Filtering.}  
For each prompt $x$, we generate a set of $M$ candidate responses $\{Y_1, \dots, Y_M\}$ using the current GQS model. Each response $Y_i = [q_1, \dots, q_N]$ ($1 \le i \le M$) consists of a list of $N$ queries. Using the CTR model described in Section 2, we compute the total predicted click likelihood of each response:
\begin{equation}
    r(Y_i) = \sum_{j=1}^N \hat{p}(q_j),
\end{equation}
where $\hat{p}(q_j)$ is the predicted CTR of query $q_j$.
To select a \emph{preferred response} $Y^c_{ctr}$, we filter candidate responses with low semantic diversity (below a fixed threshold $\delta$) and choose the one with the highest $r(Y)$. Semantic diversity is assessed using a separately trained BERT-based model, which takes the full list of suggested queries along with the user query as input and outputs a diversity score. This model is supervised using human-annotated examples reflecting different levels of diversity.

%via max-pooling of pairwise embedding distances. This model is trained using query sets manually annotated with low/medium/high diversity labels.

For the \emph{rejected response} $Y^r_{ctr}$, we follow robust preference construction principles and select from candidates with CTR scores in the lower tail of the score distribution. Specifically, we reject candidates whose CTR falls below $\mu - 2\sigma$~\cite{negative}, where $\mu$ and $\sigma$ are computed over all $r(Y_i)$ within the candidate set. This enforces that each preference pair exhibits both semantic and reward-level separation.

\paragraph{Step 2: Weighted DPO Loss.}  
Let $\pi_\theta$ be the current generation model and $\pi_{\mathrm{ref}}$ be a frozen reference model (e.g., a snapshot before DPO training). Given a preference pair $(Y^c_{ctr}, Y^r_{ctr})$ for prompt $x$, we compute the DPO loss as:

{\tiny
\begin{equation}
\mathcal{L}_{\mathrm{DPO}}(\pi_{\theta};\pi_{\mathrm{ref}}) = 
  -\log\sigma\left(
    \beta \log\frac{\pi_{\theta}(Y^c_{\mathrm{ctr}} \mid x)}{\pi_{\mathrm{ref}}(Y^c_{\mathrm{ctr}} \mid x)} 
    - \beta \log\frac{\pi_{\theta}(Y^r_{\mathrm{ctr}} \mid x)}{\pi_{\mathrm{ref}}(Y^r_{\mathrm{ctr}} \mid x)}
  \right),
\end{equation}
}
where $\beta$ is a temperature hyperparameter, and $\sigma$ is the sigmoid function.

To reflect the \emph{relative importance} of each preference pair, we compute a sample-level weight:
\begin{equation}
\alpha = \sigma\left( \gamma \cdot (r(Y^c_{\mathrm{ctr}}) - r(Y^r_{\mathrm{ctr}})) \right),
\end{equation}
where $\gamma$ is a scaling factor. Intuitively, larger CTR gaps indicate more reliable preferences and therefore contribute more strongly to model updates.

The CTR-weighted DPO loss is then defined as:
\begin{equation}
\mathcal{L}_{\mathrm{CTR\text{-}DPO}} =
\alpha \cdot \mathcal{L}_{\mathrm{DPO}}(\pi_{\theta};\pi_{\mathrm{ref}}).
\end{equation}

\paragraph{Step 3: Diversity-Aware Preference Learning.}  
In addition to CTR-guided preference learning, we introduce diversity-aware supervision to encourage \emph{semantic diversity} in the generated query lists. Specifically, we construct auxiliary preference pairs $(Y^c_{\mathrm{div}}, Y^r_{\mathrm{div}})$ where the CTR scores are close, but their diversity scores differ significantly (i.e., $|s(Y^c_{\mathrm{div}}) - s(Y^r_{\mathrm{div}})| \ge \delta$). These pairs are used to apply a DPO-style objective that encourages preference toward the more diverse response, using the same loss structure as in the CTR-aligned case.
The final training objective combines both preference signals:
\begin{equation}
\mathcal{L}_{\mathrm{CTR\text{-}DIV}} =
\mathcal{L}_{\mathrm{CTR\text{-}DPO}} + \lambda \cdot \mathcal{L}_{\mathrm{DIV\text{-}DPO}},
\end{equation}
where $\lambda$ is a tunable coefficient balancing alignment accuracy and diversity regularization.

This framework enables structured and interpretable alignment of the generation model to user behavior, while preserving diversity and avoiding generic response collapse.

\subsection{Iterative Training with Calibrated CTR Predictor}

Collecting online user click feedback for CTR model training is often time- and resource-intensive. To improve efficiency, we aim to reuse each batch of logged feedback for multiple rounds of DPO-style fine-tuning.
However, this introduces a critical challenge: after each iteration of DPO, the distribution of the generation policy $\pi_\theta$ shifts. Consequently, the CTR model—originally trained on real click data collected from queries generated by the initial policy $\pi_{\theta^{(0)}}$—becomes misaligned with the queries generated by the updated policy $\pi_{\theta^{(t)}}$. This distribution mismatch stems from a distribution drift in the CTR model, caused by a signal shift from real user clicks to predicted clicks, introducing bias into preference modeling.

To address this issue, we propose an iterative framework where a CTR model is explicitly calibrated via \textit{importance weighting} using the likelihood ratio between the initial policy and the optimized $t$ step's policy $\pi_{\theta^{(t)}}$. 

Let $\pi_{\theta^{(0)}}$ denote the generation policy used to generate the response list when the CTR training data was collected, and $\pi_{\theta^{(t)}}$ denote the policy at iteration $t$. For each response $Y = [q_1, \dots, q_N]$ for each prompt $x$ in the CTR dataset $\mathcal{D}_{\text{ctr}}$, we estimate its probability under both $\pi_{\theta^{(0)}}$ and $\pi_{\theta^{(t)}}$:

\begin{equation}
w^{(t)}(Y)=\mathrm{Clip}(\frac{\pi_{\theta^{(t)}}(Y \mid x)}{\pi_{\theta^{(0)}}(Y \mid x)}, 1 - \epsilon, 1 + \epsilon),
\end{equation}
where $w^{(t)}(Y)$ is the importance weight used to reweight the loss contribution of generated response $Y$ in the CTR training process.

We then train the CTR predictor $\hat{p}_{\phi}^{(t)}$ at iteration $t$ using a weighted binary cross-entropy loss:

{\scriptsize
\begin{equation}
\begin{aligned}
\mathcal{L}_{\mathrm{CTR}}^{(t)} = 
- \textstyle\sum_{(x, Y) \in \mathcal{D}_{\text{ctr}}} 
\textstyle\sum_{(q, y) \in Y} 
w^{(t)}(Y) \cdot \left[
\right. \\
\left.
y \log \hat{p}_{\phi}^{(t)}(q \mid x) + (1 - y) \log \left(1 - \hat{p}_{\phi}^{(t)}(q \mid x)\right)
\right],
\end{aligned}
\label{ctr_loss}
\end{equation}
}
% {\scriptsize
% \begin{equation}
% \begin{aligned}
% \mathcal{L}_{\mathrm{CTR}}^{(t)} = 
% - \textstyle\sum_{(x, q, y) \in \mathcal{D}_{\text{ctr}}} w^{(t)}(Y) \cdot \Big[
% &\, y \log \hat{p}_{\phi}^{(t)}(q \mid x) \\
% &+ (1 - y) \log (1 - \hat{p}_{\phi}^{(t)}(q \mid x)) \Big],
% \end{aligned}
% \label{contra_loss}
% \end{equation}
% }
where $y \in \{0,1\}$ denotes the click label, and $\hat{p}_{\phi}^{(t)}$ is the calibrated CTR model trained for the current policy.

Once $\hat{p}_{\phi}^{(t)}$ is trained, we regenerate the reward estimates $r(Y)$ for candidate responses, re-sample updated preference pairs $(Y^c, Y^r)$ as described in Section~\ref{sec:CTR-DrivenPO}, and perform the next round of DPO optimization over the updated  $\pi_{\theta^{(t)}}$:
% Once $\hat{p}_{\phi}^{(t)}$ is trained, we regenerate the reward estimates $r(Y)$ for candidate responses, re-sample updated preference pairs $(Y^c, Y^r)$, and perform the next round of DPO optimization over the updated policy $\pi_{\theta^{(t)}}$:
\begin{equation}
\theta^{(t+1)} \leftarrow \arg \min_{\theta} \; \mathcal{L}_{\mathrm{CTR\text{-}DIV}}^{(t)}(\pi_\theta; \pi_{\theta^{(t)}}).
\end{equation}
% where $\mathcal{L}_{\mathrm{CTR\text{-}DIV}}^{(t)}$ uses reward scores predicted by $\hat{p}_{\phi}^{(t)}$ to determine preference weights and pairwise losses, as described in Section 3.

This iterative calibration mechanism enables effective reuse of collected CTR data across multiple DPO updates while mitigating reward drift caused by evolving generation policies.

\section{Experiments}
In this section, we perform a series of experiments to evaluate the effectiveness and advantages of our proposed method.
% Specifically, we aim to address the following research questions (\textbf{RQs}): 
%分为五个问题，
%1: 设计GQS在不同任务以及不同指标上的性能
%2: 消融实验
%3: 线上效果
%4: 敏感性分析
\begin{comment}

\begin{itemize}
    \item \textbf{RQ1:} How does the proposed GQS perform across different tasks and evaluation metrics?
    \item \textbf{RQ2:} What are the impacts of the components of GQS to enhance the model performance? 
    \item \textbf{RQ3:} Does the proposed framework perform well in online service?
    \item \textbf{RQ4:} What is the effect of hyper-parameters on the efficacy of GQS?
\end{itemize}
\end{comment}

% \textbf{RQ1}: How does Baidu compare with other leading deep multi-view clustering techniques in terms of performance? \textbf{RQ2}: What art the impacts of the components of AIRMVC to enhance multi-view clustering results? \textbf{RQ3}: What clustering structures are identified by AIRMVC?

\subsection{Experimental Setup}
% \textbf{(T3)} Search-box hinting.
\paragraph{Tasks and Datasets.}
We evaluate on two query suggestion tasks in Baidu’s
% \footnote{To preserve anonymity during double-blind review, we anonymize the organization name. Details will be disclosed in the camera-ready version.}
conversational AI assistant, with details shown in the Appendix \ref{sec:task}:
\textbf{(T1)} General query suggestion and 
\textbf{(T2)} Creative-writing snippet generation.  Here, we adopt real large-scale search logs from the Baidu AI assistant for model training and evaluation.
Specifically, for each task, we first construct a dataset for CTR predictor training via sampling 14 days of click logs from the initial SFT baseline. Afterwards, we randomly sample 10,000 queries from the 15th-21st days of search logs as the training set for CTR alignment, with 10,000 queries from the 22nd day as the test set.

\paragraph{Evaluation Metrics and Baselines.}
To provide a thorough evaluation of the model's performance, we utilize three metrics: Click-Through Rate (CTR), Relevance (Rel.) and Diversity (Div.). Besides, we adopt ERNIE Speed (21B) \cite{sun2020ernie} as our foundation model. We evaluate our approach in comparison with four baselines: SFT, KTO~\cite{ethayarajh2024_KTO}, SimPO~\cite{meng2024simposimplepreferenceoptimization}, and DPO~\cite{NEURIPS2023_DPO}, where preference pairs are derived from (1) user click behaviour and (2) estimated CTR values.
The AI assistant presents 3 and 8 query suggestions as clickable items for Task 1 and Task 2, respectively. Besides, we set $\lambda$ as 0.1 for all experiments.

\begin{table}[]
\centering
\resizebox{0.99\linewidth}{!}{
\begin{tabular}{c c ccc ccc}
\toprule
\multirow{2}{*}{Scenario}      & \multirow{2}{*}{Method} & \multicolumn{3}{c}{Task1}            & \multicolumn{3}{c}{Task2}             \\
\cmidrule(lr){3-5} \cmidrule(lr){6-8}
                               &                         & CTR Impr. & Rel. & Div. & CTR Impr. & Rel. & Div. \\
\midrule
\multirow{2}{*}{Base}          & SFT                     & 0.00        & 80.53       & 85.63     & 0.00        & 84.86       & 81.42     \\
                               & Few-shot                & 1.14        & 78.12       & 80.17     & -5.81       & 82.66       & 79.46     \\
\midrule
\multirow{4}{*}{\makecell{Click-\\Aligned}}   & SFT$_\text{clk}$        & 4.61        & 81.89       & 83.49     & 0.76        & 84.14       & 78.11     \\
                               & KTO$_\text{clk}$        & -1.61       & 79.45       & \textbf{86.81}     & -1.69       & 84.55       & 81.97     \\
                               & SimPO$_\text{clk}$      & -1.71       & 78.34       & 84.36     & 1.91        & 86.05       & 78.20     \\
                               & DPO$_\text{clk}$        & -1.42       & 80.91       & 83.94     & 1.54        & 84.65       & 76.78     \\
\midrule
\multirow{4}{*}{\makecell{CTR-\\Aligned}}     & SFT$_\text{ctr}$        & 19.44       & 84.08       & 80.12     & 4.45        & 87.32       & 75.14     \\
                               & KTO$_\text{ctr}$        & 30.78       & 86.74       & 77.68     & 7.63        & 89.04       & 72.12     \\
                               & SimPO$_\text{ctr}$      & 32.99       & 90.36       & 79.29     & 16.98       & 93.86       & 70.61     \\
                               & DPO$_\text{ctr}$        & 60.15       & 91.15       & 65.73     & 25.51       & 95.83       & 59.43     \\
\midrule
Ours                           & GQS                    & \textbf{70.36}       & \textbf{94.60}       & 86.04     & \textbf{30.72}       & \textbf{98.11}       & \textbf{82.09}     \\
\bottomrule
\end{tabular}}
\caption{The overall performance of various models for GQS. Results of three categories of methods are presented: 1) \textbf{Base} means the model is trained by original user query log without click alignment, 2) \textbf{Click-Aligned} denotes the model is trained by click/non-click logs as chosen/rejected instances, without any CTR-based calibration, 3) \textbf{CTR-Aligned} represents the model is aligned with predicted CTR scores. ``Impr.'' denotes the relative improvement (\%), ``Rel.'' denotes semantic relevance, and ``Div.'' denotes semantic diversity.}
\label{com_res}
\end{table}

% \paragraph{Implementation Details.}
% We leverage 14 days of click logs from the initial SFT baseline to train our CTR model. For preference-alignment training, we randomly sample 10,000 queries from days 15–21 and evaluate on 10,000 queries from day 22, all drawn from Baidu Assistant’s production logs.

\subsection{Overall Results and Analysis}
%In this subsection, we conduct comprehensive experiments to demonstrate the superior and effectiveness of our designed GQS.  
The overall experimental results are presented in Table~\ref{com_res}, and we have the following observation:

%sft和few-shot效果差不多，使用大量数据进行sft，并不一定能提高模型的效果。说明指令遵循能力的强弱对于生成式推荐任务来说并没有决定性的作用
%整体来看，基于点击信息对LLM进行偏好对齐能够提升模型生成RQs的点击率。例如直接用点击样本进行sft就能够略微提升ctr。然而，直接基于点击/未点击样本来构造正负样本的效果并不能得到保证，例如Baidu,显著降低了效果。这是因为用户的点击信号和未点击信号都非常noisy和random，损害了偏好学习的效果
%基于预估ctr信号进行偏好对齐的方法，能够获得稳定的提升。具体来说，只使用高ctr筛选的正样本，就能获得显著的提升。说明我们基于ctr的筛选方式能够获得高质量的高ctr样本。当加上低ctr的rejected resposne作为负样本时，kto, simpo，dpo都有显著提升，说明偏好对齐相比sft更有效果，其中，基于pair样本的方法如simpo和dpo比kto效果好，说明pair样本更容易使模型理解用户偏好。 Iter DPO效果好于DPO，说明多次迭代能够进一步提高模型的预估ctr。
%\begin{itemize}
    In both the Click-Aligned and CTR-Aligned scenarios, we observe that improvements in CTR performance are accompanied by enhanced relevance but reduced diversity. This trade-off arises because high-CTR samples often share similar patterns, which biases the model toward generating less diverse suggestions.

    For Click-Aligned, although real user click data were used for preference alignment, we observe an overall decline in CTR in both Task 1 and Task 2. For example, in Task 1, applying the KTO, SimPO, and DPO algorithms led to CTR decreases of 1.61\%, 1.71\%, and 1.42\%, respectively. We attribute this performance drop to the noise and randomness inherent in users’ click and non-click signals, which undermine the effectiveness of preference learning and highlight the limitations of directly using raw click data.

    In comparison, methods that perform preference alignment based on predicted CTR signals consistently yield improvements. This indicates that filtering based on CTR scores predicted by CTR models is more significant than direct filtering based on click signals, thereby reducing noise and variance. This demonstrates that preference alignment is more effective than SFT alone, and understanding the patterns of negative samples is crucial for the model. 

    Our proposed GQS method jointly optimizes CTR‐prediction preferences and diversity, thereby achieving CTR gains while simultaneously preserving high diversity and relevance. For example, in Task 2, compared to the DPO algorithm in the CTR-Aligned scenario, GQS improves CTR, relevance, and diversity by 5.21, 2.28, and 22.66 points, respectively.

    % \item The performance of SFT with human-annotated training data and in-context learning using few-shot examples is similar in Task 1, suggesting that utilizing more annotated data for SFT does not necessarily enhance recommendation performance. This implies that the strength of instruction-following capabilities is not a decisive factor for generative recommendation tasks. 
    
    % This is reasonable because even with high-quality annotations for GQS tasks, there remains a significant discrepancy between the annotators' preferences and the users' actual click preferences.    
    % Overall, aligning preference directly with click samples can moderately improve the AEC of the generated recommendation queries, although the effectiveness is not always guaranteed. For instance, using click samples for SFT can slightly improve ACC from 0.1233 to 0.1310 in Task 1, representing a $6.24\%$ increase. However, constructing positive and negative samples for preference alignment based solely on clicked/non-clicked samples might degrade performance, as seen in the case of KTO\(_{clk}\)  (-1.87\%), SimPO\(_{clk}\)(-1.30\%), DPO\(_{clk}\)(-0.97\%) for Task 1, and KTO\(_{clk}\) (-1.49\%) for Task 2. This is due to the noisy and random nature of user click and non-click signals, which adversely affects preference learning. 
    
%\end{itemize}

\begin{figure}[t]
    \centering
    \includegraphics[width=0.99\linewidth]{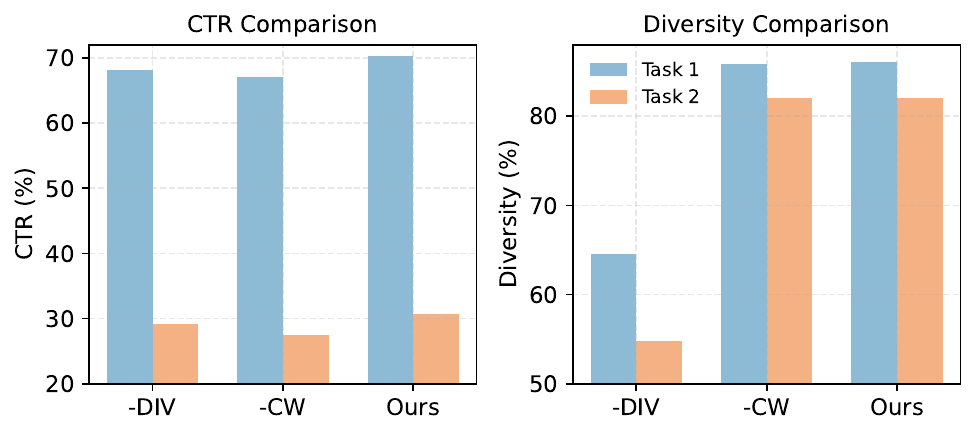}
    \caption{Ablation results of Preference Alignment}
    \label{ab_res}
\end{figure}

\begin{figure}[t]
    \centering
    \includegraphics[width=0.99\linewidth]{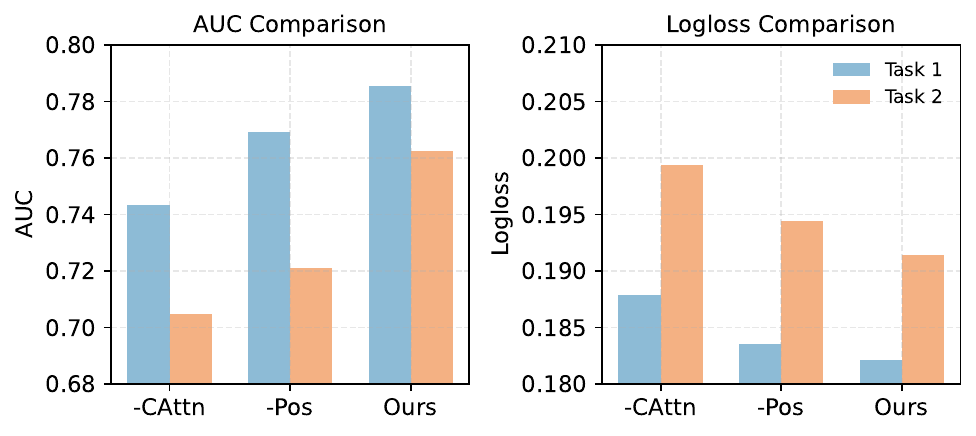}
    \caption{CTR model performance (AUC and Logloss) under different architectural components.}
    \label{ab_ctr}
\end{figure}

\subsection{Ablation Studies}
In this subsection, we present experiments to evaluate the effectiveness of key components of GQS. %We first conduct ablation studies to disentangle the impact of major modules in the preference alignment strategy, including the diversity-aware learning objective and the CTR-weighting mechanism. Additionally, we perform a structural ablation on the multi-source CTR prediction model to assess the contributions of its architectural components.

\noindent \textbf{Ablation of preference alignment modules.}
We denote ``-DIV'' as the removal of the diversity-aware learning component, and ``-CW'' as the exclusion of the CTR-weighting strategy. As illustrated in Figure~\ref{ab_res}, ablating ``-CW'' leads to a substantial drop in CTR, highlighting the value of emphasizing preference pairs with larger CTR gaps. Removing ``-DIV'' results in a notable decrease in diversity scores, underscoring the role of diversity regularization in enhancing generative variability. Interestingly, CTR performance also slightly declines without the diversity objective, suggesting that diverse suggestions are more likely to capture user interest.

\noindent \textbf{Analysis of CTR Model.}
Here, ``-CAttn'' refers to the removal of the cross-attention mechanism for multi-source fusion; in this variant, all contextual inputs are simply concatenated and fed into a BERT encoder. ``-Pos'' denotes the exclusion of positional embeddings. As shown in Figure~\ref{ab_ctr}, removing either component leads to a drop in AUC and an increase in Logloss, indicating degraded prediction quality. These results highlight the importance of cross-attention in effectively integrating heterogeneous information, as well as the crucial role of positional embeddings in modeling position bias.
%Due to the space limtited, we present additional experiments on Sec.~\ref{add_res_ab} of Appendix.
\subsection{Effect of Iterative Training with CTR Calibration}
\begin{figure}[t]
  \centering
  \includegraphics[width=0.99\linewidth]{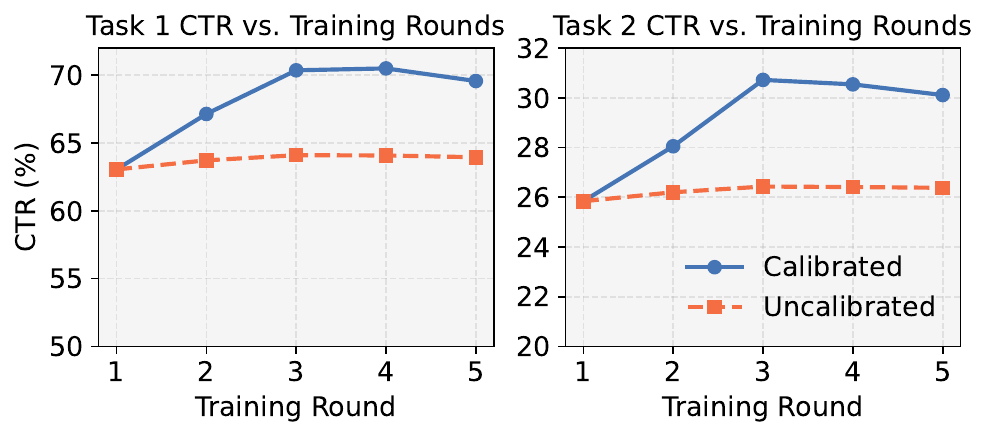}
  \caption{CTR progression over training rounds in Task 1 and Task 2, comparing calibrated vs. uncalibrated preference guidance.}
  \label{fig:ctr_iterative_comparison}
\end{figure}

We evaluate the impact of our iterative training strategy, where the CTR model is progressively updated to guide preference optimization. As shown in Figure~\ref{fig:ctr_iterative_comparison}, CTR performance improves notably in the first few rounds—for example, Task~1 improves by 7.45 points (from 63.05 to 70.50)—and stabilizes afterward. This demonstrates that iterative calibration effectively enhances alignment quality without overfitting.
In contrast, a baseline using a fixed CTR model across rounds yields only marginal improvements (e.g., Task~1 increases by 1.03 points, from 63.05 to 64.08), indicating the limited reliability of static preference signals. These results highlight the benefit of updating the reward model to reflect evolving generation behavior.
We also observe a saturation trend beyond the third round, where gains become negligible or slightly decline (e.g., Task~2 drops by 0.61 points, from 30.72 to 30.11). This suggests that excessive iterations may offer diminishing returns, reinforcing the need to select a moderate number of training rounds to balance effectiveness and efficiency.

\section{Conclusion}

We propose \textbf{GQS}, a generative query suggestion framework for conversational search that aligns with user preferences through multi-source CTR modeling, diversity-aware preference optimization, and iterative CTR calibration. Extensive experiments on real-world tasks demonstrate that GQS consistently improves CTR, relevance, and diversity. Our findings highlight the importance of modeling user feedback and maintaining semantic diversity in preference-aligned generation.

\newpage

\section*{Limitations}
While GQS demonstrates strong performance in aligning query suggestions with user preferences, several limitations remain. First, our approach relies on click-through data, which may carry inherent biases such as position effects and delayed feedback. Although the CTR model helps mitigate this, residual bias could still influence optimization. Second, the diversity estimation module is trained on a limited number of annotated samples, which may not generalize well to unseen topics or domains. Third, the iterative optimization process requires multiple rounds of reward model updates and generation training, which increases computational cost. In future work, we plan to explore more efficient update strategies and incorporate human-in-the-loop feedback for better alignment quality.

\section*{Ethics Statement}
\paragraph{Use of AI Assistants}
We have employed ChatGPT as a writing assistant, primarily for polishing the text after the initial composition.

% \section*{Acknowledgments}

% This document has been adapted
% by Steven Bethard, Ryan Cotterell and Rui Yan
% from the instructions for earlier ACL and NAACL proceedings, including those for
% ACL 2019 by Douwe Kiela and Ivan Vuli\'{c},
% NAACL 2019 by Stephanie Lukin and Alla Roskovskaya,
% ACL 2018 by Shay Cohen, Kevin Gimpel, and Wei Lu,
% NAACL 2018 by Margaret Mitchell and Stephanie Lukin,
% Bib\TeX{} suggestions for (NA)ACL 2017/2018 from Jason Eisner,
% ACL 2017 by Dan Gildea and Min-Yen Kan,
% NAACL 2017 by Margaret Mitchell,
% ACL 2012 by Maggie Li and Michael White,
% ACL 2010 by Jing-Shin Chang and Philipp Koehn,
% ACL 2008 by Johanna D. Moore, Simone Teufel, James Allan, and Sadaoki Furui,
% ACL 2005 by Hwee Tou Ng and Kemal Oflazer,
% ACL 2002 by Eugene Charniak and Dekang Lin,
% and earlier ACL and EACL formats written by several people, including
% John Chen, Henry S. Thompson and Donald Walker.
% Additional elements were taken from the formatting instructions of the \emph{International Joint Conference on Artificial Intelligence} and the \emph{Conference on Computer Vision and Pattern Recognition}.

\bibliography{custom}

\appendix

\onecolumn
\section{Prompt Construction and COO Information Refilling}
\subsection{Prompt and Respones Format}
\label{sec:prompt}
Here we provide the detailed prompt and response format used for query generation.
\begin{figure*}[h]
\begin{tcolorbox}[
    title=Example Prompt and Response for Query Generation,
    colframe=colframecolor,
    colback=colbackcolor,
    label={box:gqr_prompt_example}
]
\textbf{Instruction:}

You are an intelligent assistant helping users explore information in a conversational search session.

\textbf{[Current User Query]:} \\
\textit{How does intermittent fasting affect metabolism?}

\textbf{[Current Assistant Response]:} \\
\textit{Intermittent fasting can influence metabolic health by...}

\textbf{[Conversation History Before This Turn] ($h_t$):} \\
\textit{User: What are effective diet methods for weight loss? \\
Assistant: There are various methods such as calorie restriction, low-carb diets, and intermittent fasting...}

\textbf{[User Profile Features]:} \\
\textit{Age: 35; Interests: fitness, healthy diet, sustainable health practices}

\textbf{[Co-occurring Queries]:} \\
\textit{Benefits of fasting for fat loss; Best time windows for intermittent fasting; Intermittent fasting vs. calorie restriction}

Based on the current query and answer, previous conversation history, user profile, and related queries, generate 3 new and diverse follow-up queries that the user may find useful.

\hrulefill

\textbf{Generated Response:}

Does fasting improve insulin sensitivity? \\
Best fasting methods for beginners \\
Fasting and muscle preservation tips 
\end{tcolorbox}
\end{figure*}

%\begin{comment}
\subsection{Co-Occurrence Query Information Construction and Refilling}
\label{sec:coo_con}
\begin{figure*}[h]
    \centering
    \includegraphics[width=0.9\linewidth]{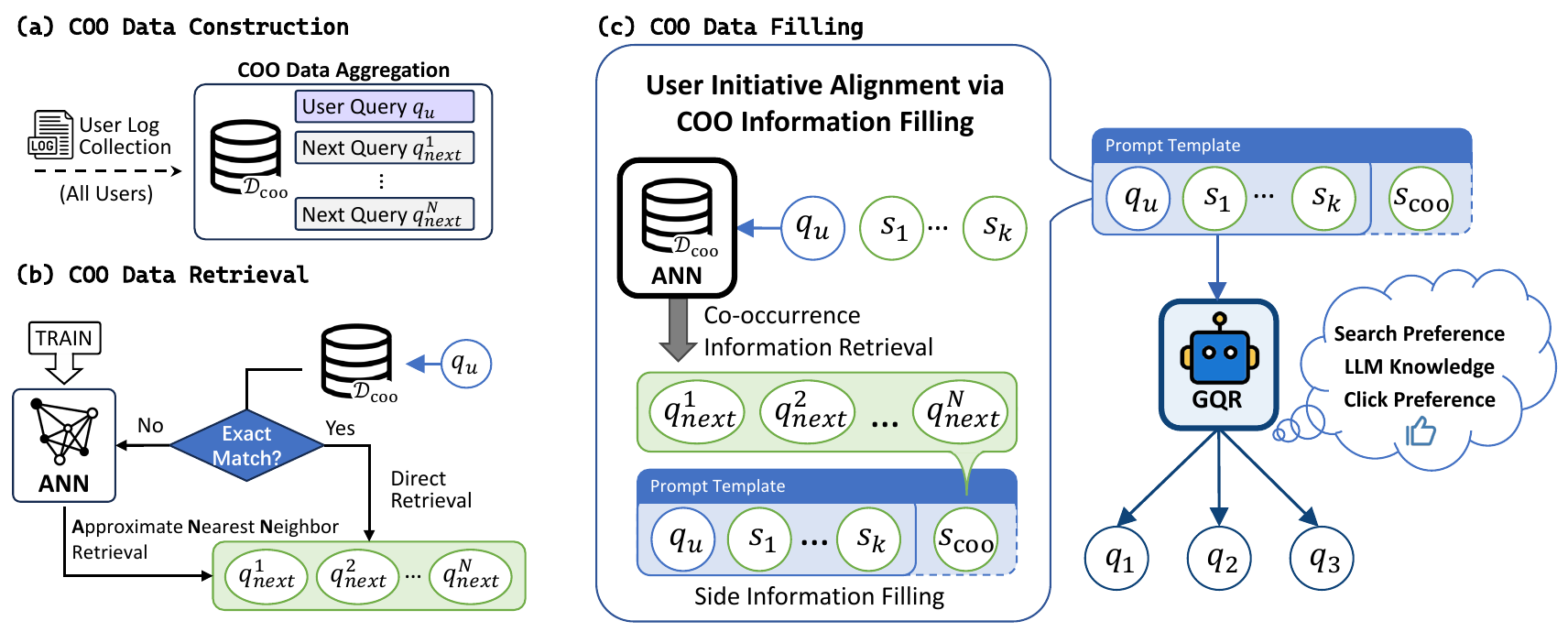} % 
    \caption{COO query information construction and refilling as side information.}
    \label{fig:coo}
\end{figure*}

We demonstrate how co-occurring query information is constructed and applied in query suggestion, as illustrated in Figure~\ref{fig:coo}.
% This template frames diverse query-centric tasks (e.g., query prediction, query suggestion) as a generation process.
\twocolumn

As shown in Figure~\ref{fig:coo}, given an input query $q$ and optional side information $S = {s_1, s_2, \ldots, s_k}$, the LLM $\mathcal{M}$ generates suggested queries $RQ = {q_1, q_2, \ldots, q_N}$. The side information $S$ is task-specific, such as session history or system responses in conversational search as shown in Section \ref{sec:prompt}. 
To incorporate co-occurring query signals as side information for aligning with user search preferences, we construct COO information as follows:
(a) We extract co-occurrence (COO) signals $(q_u, q_{\text{next}}, c)$ from Baidu’s complete user query logs, where $q_u$ is the user query, $q_{\text{next}}$ is a co-occurring query in the same session, and $c$ denotes the frequency of this COO. These signals are aggregated into a COO dictionary $D_{\text{COO}}$.
(b) We build an Approximate Nearest Neighbor (ANN) semantic matching system over $D_{\text{COO}}$, enabling efficient retrieval of relevant co-occurring queries based on user input. If the input query $q_u$ has an exact match in $D_{\text{COO}}$, we retrieve its corresponding co-occurring queries; otherwise, we retrieve semantically similar queries from the ANN index to provide meaningful co-occurring suggestions.
(c) The retrieved co-occurring queries are formatted as textual side information and incorporated into the prompt template to help align generated suggestions with user search preferences.
%\end{comment}

\begin{comment}
\begin{figure}[t]
\centering
\begin{minipage}{0.45\linewidth}
\centerline{\includegraphics[width=\textwidth]{Fig/ab/Task1_ab_Div.png}}
\vspace{3pt}
{\centerline{Task1-Div}}
\end{minipage}
\hspace{5pt}
\begin{minipage}{0.45\linewidth}
\centerline{\includegraphics[width=\textwidth]{Fig/ab/Task2_ab_Div.png}}
\vspace{3pt}
{\centerline{Task2-Div}}
\end{minipage}
\caption{Ablation studies of proposed modules.}
\label{ab_res_div}
\end{figure}

\section{Addtional Ablation Stuides.}\label{add_res_ab}

Due to page limitations, we supplement this section with additional ablation results, covering both the ablated modules and the CTR model architecture. As shown in Figure~\ref{ab_res_div}, removing the diversity‐pair construction causes a steep decline in diversity metrics, owing to the loss of the diversity constraint. Furthermore, Figure~\ref{ab_CTR_add} demonstrates that each design component of the CTR model contributes significantly to its overall performance.

\begin{figure}[t]
\centering
\begin{minipage}{0.45\linewidth}
\centerline{\includegraphics[width=\textwidth]{Fig/ab/Task1_logloss_ab.png}}
\vspace{3pt}
{\centerline{Task1-Logloss}}
\end{minipage}
\hspace{5pt}
\begin{minipage}{0.45\linewidth}
\centerline{\includegraphics[width=\textwidth]{Fig/ab/Task2_AUC_ab.png}}
\vspace{3pt}
{\centerline{Task2-AUC}}
\end{minipage}
\caption{Ablation studies of CTR structure.}
\label{ab_CTR_add}
\end{figure}

\end{comment}

\section{Evaluation Criteria}
\label{criteria}
\textbf{Relevance.}  
This metric evaluates the semantic alignment between a suggested query and the user's original query. Relevance is categorized into three levels:  
\begin{itemize}
    \item \textit{High (1.0):} The suggested query is highly related to the user's query in both topic and intent, providing additional and pertinent information that complements the original query.
    \item \textit{Moderate (0.5):} The suggested query is somewhat related but may not directly address the user's core intent, often offering tangential or loosely associated content.
    \item \textit{Low (0.0):} The suggested query shows minimal topical or semantic overlap with the user's query, failing to align in intent or subject matter.
\end{itemize}

\textbf{Diversity.}  
This metric assesses the degree of novelty and non-redundancy of each suggested query along three dimensions:  
\begin{itemize}
    \item \textit{Exclusivity from the user query:} The suggested query should not be semantically equivalent to the original query, ensuring it introduces a new perspective or subtopic.
    \item \textit{Exclusivity from the AI response:} The suggested query should elicit new content from the AI model rather than reproducing information already covered in the initial response.
    \item \textit{Exclusivity from other suggested queries:} Each suggested query in the recommended set should be distinct from others, contributing unique value to the overall list.
\end{itemize}
Scoring rule: A diversity score of 1.0 is assigned if all three aspects are satisfied, 0.5 if exactly two are satisfied, and 0.0 otherwise.

We utilize GPT-4o as an automatic judge to score both relevance and diversity. Prompts are carefully designed to instruct the model to assess each dimension independently and output scores following the defined criteria.

\section{Sensitive Analysis}\label{rq4}
In this subsection, we conduct experiments on Task~1 to analysis the sensitivity of the hyper-parameters in this paper.

\begin{figure}[t]
    \centering
    \includegraphics[width=0.45\textwidth]{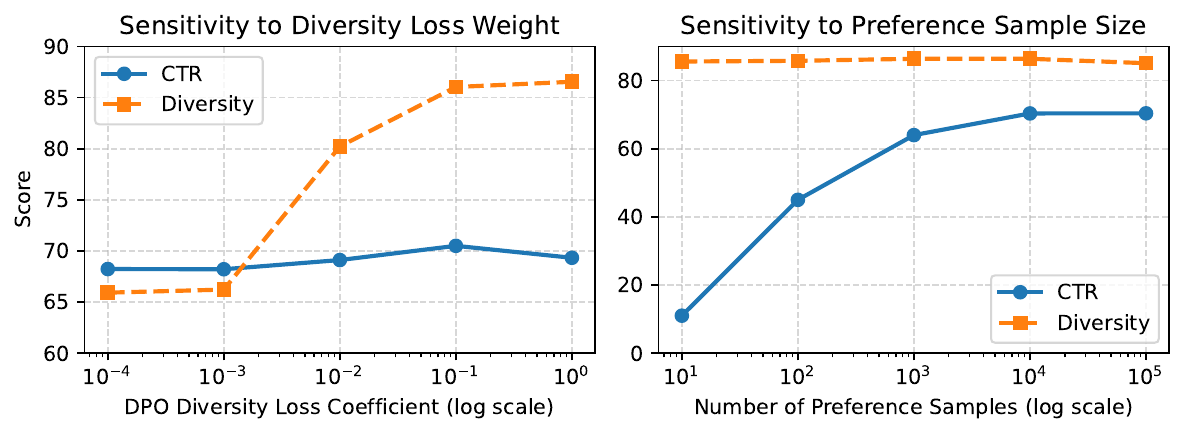}
    \caption{Sensitivity analysis results. (a) Effect of diversity loss coefficient $\lambda$ on CTR and diversity. (b) Effect of alignment sample size on model performance.}
    \label{sen_res}
\end{figure}

\paragraph{Sensitivity Analysis of Diversity Loss Coefficient $\lambda$.}  
We conduct experiments to analyze the impact of the diversity loss weight $\lambda$ in our preference optimization. We test values from $\{1.0, 0.1, 0.01, 0.001, 0.0001\}$ and summarize the results in Figure~\ref{sen_res}(a). The results show that moderately weighted diversity loss leads to better balance between CTR and diversity. For instance, $\lambda = 0.1$ achieves the best CTR (70.50) while maintaining high diversity (86.04). When $\lambda$ becomes too small, the diversity benefit diminishes significantly (e.g., 66.22 at $\lambda=0.001$), though CTR still remains relatively stable. Conversely, larger values like $\lambda=1.0$ ensure high diversity (86.55), but lead to a lower CTR (69.33). These results highlight the importance of moderate regularization strength—$\lambda=0.1$ offers the best overall trade-off.

\paragraph{Sensitivity Analysis of Number of Alignment Samples.}  
We further examine the effect of the number of preference samples used in alignment training. We vary the sample count in $\{10, 10^2, 10^3, 10^4, 10^5\}$ and report the performance on Task 1 in Figure~\ref{sen_res}(b). The results show that increasing the number of samples significantly improves CTR and diversity up to a certain threshold. For example, when using only 10 samples, CTR is 11.0 and diversity is 85.6, while using 1000 samples leads to 64.0 CTR and 86.4 diversity. The model saturates around $10^4$ samples, with CTR reaching 70.36 and diversity maintaining 86.4. Further increasing the sample size to $10^5$ yields marginal gains in CTR (70.39) but causes a slight drop in diversity (85.1). Therefore, we adopt $10^4$ samples in our default setup, balancing computational cost and performance.

\section{Task Showcase}
\label{sec:task}

\begin{figure}[t]
\centering
\includegraphics[width=0.49\linewidth]{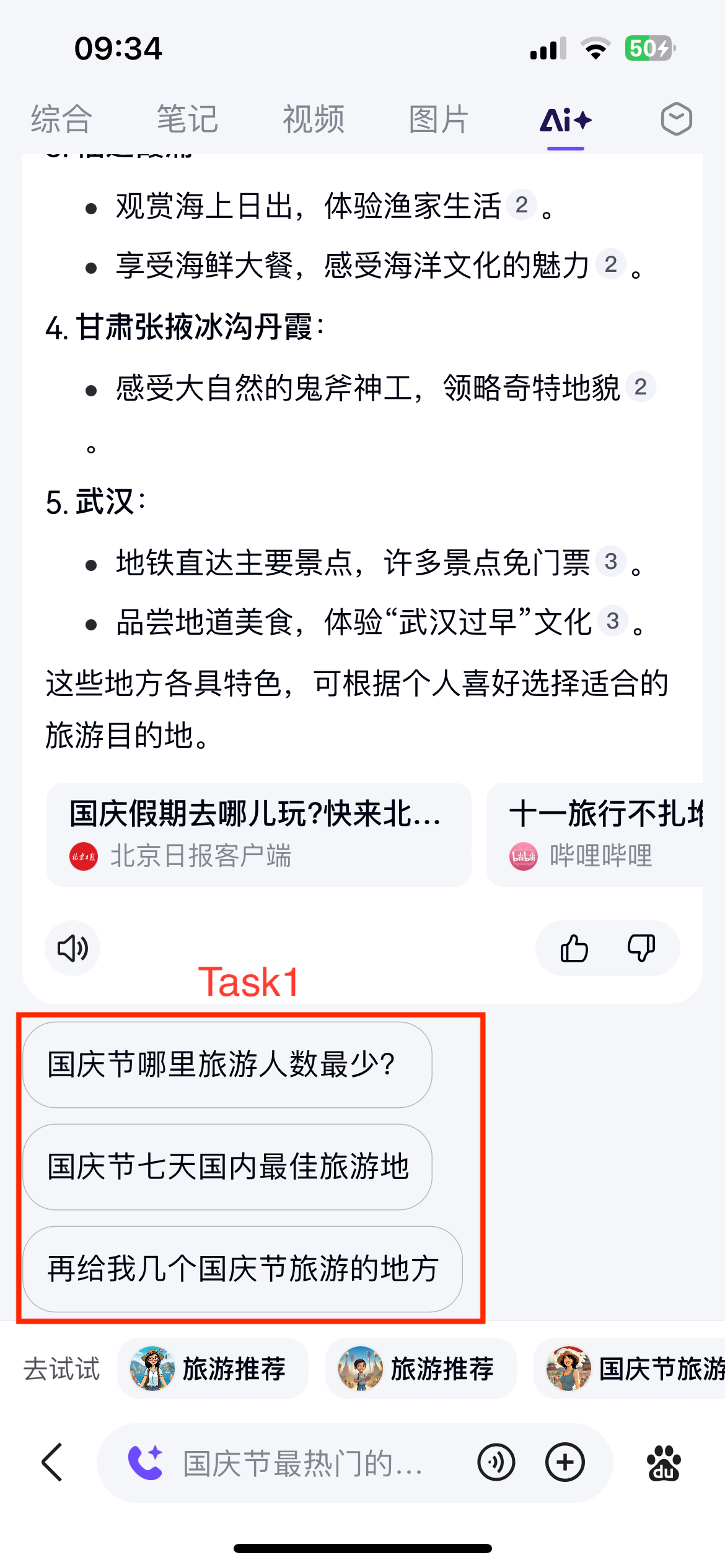} 
\includegraphics[width=0.49\linewidth]{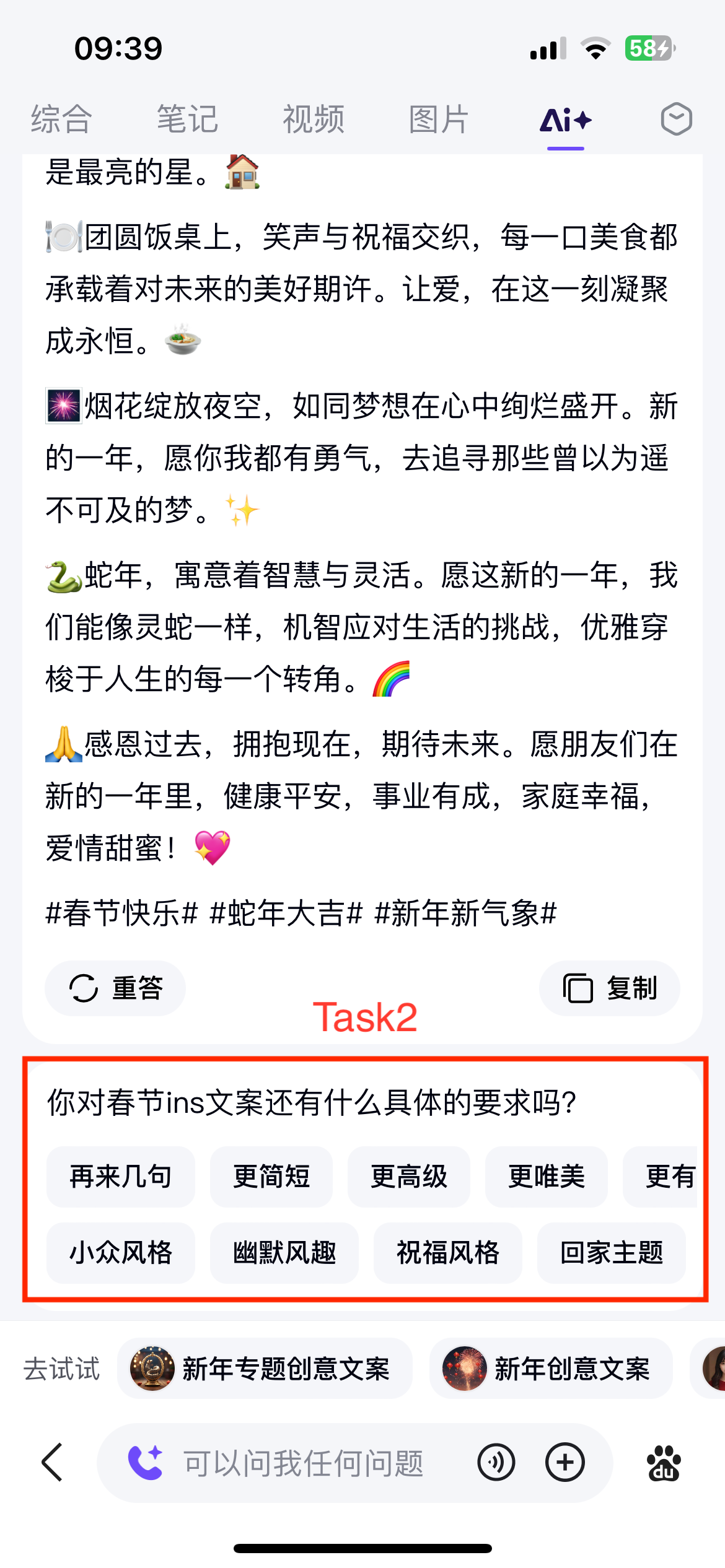} 
\caption{Illustration of the three query suggestion tasks in the Baidu Conversational Search System.}
\label{fig:showcases}
\end{figure}

\paragraph{Tasks and Datasets.}
We evaluate our framework on two query suggestion tasks in Baidu’s conversational AI assistant, with examples shown in Figure~\ref{fig:showcases}:

\textbf{(T1) General query suggestion.} After answering a user’s query, the system offers three follow-up query suggestions for preference elicitation.

\textbf{(T2) Creative-writing snippet generation.} Users receive a guiding sentence and approximately eight clickable snippet suggestions (e.g., “shorter”, “more creative”, “more elegant”) for refining creative-writing prompts.

% \textbf{(T3) Search-box hinting.} The assistant suggests a query in the search box based on the current session’s context, analogous to features in commercial search systems.

\end{document}